\documentstyle[twocolumn,pre,aps,epsf]{revtex}
\newcommand{\epsplace}[1]{\epsfxsize=3.3in \epsfbox{#1}}
\newcommand{\be}{\begin{equation}}
\newcommand{\ee}{\end{equation}}
\newcommand{\bea}{\begin{eqnarray}}
\newcommand{\eea}{\end{eqnarray}}

\newcommand{\bbmc}{\begin{multicols}{2} }
\newcommand{\eemc}{\end{multicols}}

\title{Random walks in the space of conformations of toy proteins}
\author{Rose Du,${}^{1}$ Alexander Yu. Grosberg,${}^{1,2}$ Toyoichi Tanaka${}^{1}$}

\address{
${}^1$Department of Physics and Center for Materials Science and
Engineering, \\ Massachusetts Institute of Technology, Cambridge,
Massachusetts 02139,  USA \\
${}^2${\em On leave from:\/} Institute of Chemical Physics,
Russian Academy of Sciences, Moscow 117977, Russia
}
 
\address{ {\em \bigskip \begin{quote}
Monte Carlo dynamics of the lattice 48 monomers toy protein is 
interpreted as a random walk in an abstract (discrete) space of 
conformations.  To test the geometry of this space, we examine 
the return probability $P(T)$, which is the probability 
to find the polymer in the native state after $T$ Monte Carlo 
steps, provided that it starts from the native state at the initial 
moment.  Comparing computational data with the theoretical expressions 
for $P(T)$ for random walks in a variety of different spaces, we show
that conformational spaces of polymer loops may have 
non-trivial dimensions and exhibit negative curvature 
characteristic of Lobachevskii (hyperbolic) geometry.  
\end{quote} }}

\begin{document}
\maketitle

Levinthal's paradox \cite{Levinthal} 
is universally considered to be the essence of the 
protein folding problem.  In its most direct form, the paradox 
revolves around the exponentially large number of possible 
conformations being immeasurably larger than what a protein can 
conceivably test within the observable time scale.  
Levinthal's paradox arises from thinking of protein folding as
a search in a conformational space resembling a golf course 
with just one hole representing the native state.  To resolve 
the problem, it has been conjectured in the literature 
\cite{Fersht,Onuchic,Wolynes,Shakhnovich} that volume interactions
between monomers should provide an energetic bias towards the native state.  
However undoubtedly correct, this should not overshadow the 
necessity to understand the search in conformational space.  
Indeed, if we start from an open coil conformation, then 
volume interactions cannot provide any significant bias for 
a while, at least until after some minimal number of contacts 
has been formed.  In macroscopic terms, this initial stage 
is an uphill climb over an entropic barrier.  In microscopic 
terms, it is a random walk in conformation space.
In order to initiate the energy-driven downhill slide 
towards the native state, or to enter the funnel-shaped 
\cite{Onuchic,Wolynes} area of the free energy landscape 
\cite{landscape_note}, 
a fluctuation has to provide a sufficient decrease in entropy, 
or, in other words, a random walk has to bring the 
system into the specific region in conformation space.  

This way of thinking implies 
the following resolution of Levinthal's paradox:  there is no 
need for a random unbiased search to detect a single native 
state, it only needs to bring the system into some region,  
$\omega$, in the  
conformation space.  Physically, $\omega$ corresponds
to the transition (macro)state, most likely to a critical 
nucleus of some kind \cite{Fersht,Shakhnovich_nucleus,Thirumalai_nucleus}.  
Therefore, the volume 
and shape of $\omega$ in conformation space are dictated by both sequence 
specific energy factors and sequence independent properties 
of conformations.  The system searches for this critical 
(macro)state $\omega$ through a random walk in conformation space, 
largely unaffected by heteropolymeric interaction energies.  
Understanding this process is the problem of normal 
polymer dynamics \cite{Doi-Edwards} and in some cases it 
may be reduced to the kinetics of a homopolymer collapse 
\cite{Homopolymer_collapse}. Unfortunately, this problem 
is rather difficult and remains out of reach of current 
simulation techniques. 

In order to pave the way to it, we 
will consider in this work another related problem of 
random walks in conformation space, namely  
that of fluctuations around the native state.  This is itself a 
pressing issue in protein folding theory.  Indeed, 
understanding these fluctuations is necessary in order to address 
the corrections to mean field theory 
which is formulated in terms of the Random Energy Model 
(see \cite{Sfatos} and the review \cite{RMP} with 
a multitude of references therein).

\begin{figure}
\centerline{\epsplace{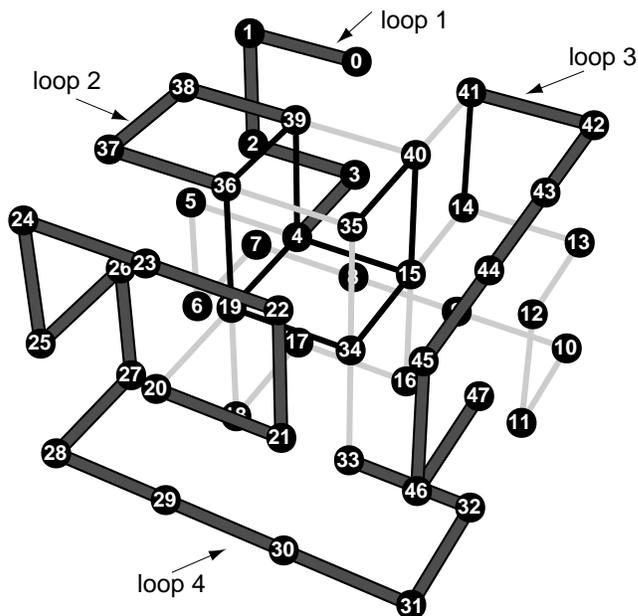}}
\caption{The 48-mer conformation.  The dark contacts 
indicate critical folding nucleus for this particular 
conformation, according to the data of the work by Mirny et al
\protect\cite{Mirny}. 
The loops considered in the present work are shown as thick lines. 
\label{fig:loops}}
\end{figure}

We will restrict ourselves with the standard lattice
protein model of 48 monomers.  In particular, we choose 
to work with the particular native state conformation 
addressed in \cite{Mirny}.  In order to 
remain in the vicinity of the native state, we can 
permanently fix the contacts which form the 
nucleus.  The nucleus conformation, as conjectured in 
\cite{Mirny}, is shown in Figure~\ref{fig:loops}.  
Finding the nucleus even for one given native conformation is
a difficult and unresolved problem in protein folding.
Mirny et al. obtained the nucleus for the native state shown in Figure~\ref{fig:loops}
by a procedure in which interactions between various
monomer pairs were mutated.
Those interactions which are necessary for a stable folded configuration 
were examined for a conserved set.  The contacts belonging to the conserved set
were inferred to be the nuclear contacts (see Figure~\ref{fig:loops}).  
As a matter of fact, the procedure for determining the nucleus 
remains a subject of scrutiny and heated debate 
\cite{nucleus_debate_Wolynes,nucleus_debate_Shakhnovich,nucleus_debate_Thirumalai,PGT_current_opinion,Pande_Rokhsar}.  
The various models of a single nucleus \cite{Shakhnovich_nucleus}, of 
multiple nuclei \cite{Thirumalai_nucleus}, and of nucleation classes 
\cite{Pande_Rokhsar} are being debated 
but the choice among the different models is not the subject of this work.
The nucleus from the work \cite{Mirny} which is being used in the present work
has been chosen arbitrarily among a large number
of possible conformations with nuclei surrounded by loops.

In order to address purely entropic factors, we shall examine 
folding of the polymer
under no interactions (except for the constraints of 
polymer connectivity, excluded volume, and fixed contacts).
The resulting structure is essentially that of many loops
with fixed ends in the nucleus.  For the discrete lattice 
model, we consider the conformational phase space as a 
graph in which the conformations are represented
by nodes on the graph. If two conformations can be 
interconverted via a single Monte Carlo move (end flip, corner flip,
or crankshaft), their corresponding nodes are connected by 
edges on the graph \cite{Homo_Folding}.  
A Monte Carlo run is thus equivalent to a random walk
on that graph.  From that point of view, 
folding  is equivalent to 
performing a random walk which returns to the origin.  
Thus, our plan is as follows:  we will perform a long 
Monte Carlo run of the above described loop model, and we will 
record all the time moments (or Monte Carlo steps) of spontaneous 
folding, or random arrival to the origin, which is the native state.  
This will give us the return probability, $P(T)$, as the function of 
Monte Carlo time, $T$.  In order to interpret these data, we will compare them 
with a summary of the known results for 
the expected behavior of $P(T)$ in a variety of different spaces.
Before proceeding with this plan, two short comments must be made:
\noindent (a) We are referring to the return probability $P(T)$, not the 
first return probability.  In other words, the 
system said to return to the origin at time $T$, no matter how many 
times it may have visited origin previously.
\noindent (b) There is a potential source of terminological confusion 
due to the well known analogy between polymer conformations and 
trajectories of random walks.  Indeed, the conformation shown in 
Figure \ref{fig:loops} is often described in terms of some 
walker in $3D$ space.  We would like to stress 
that we are speaking here about a completely different random 
walk.  In our case, the walker is the entire protein chain, and 
the walk is being performed in the abstract space of conformations.  

We are now ready to begin with a summary of the known results for the
return probabilities in spaces of various geometries.  We 
will consider only discrete spaces, assuming every elementary 
step of the random walk to be of unit length.  

$\bullet$  For a random walk of $T$ steps in an unbounded 
Euclidean space of dimension $d$, 
the probability of return is
\be
P(T) = \left( 2 \pi T /d \right)^{-d/2} \sim T^{-d/2} \ .
\label{eq:Euclidean}
\ee 
A square (or cubic) lattice is in this sense the discrete counterpart of Euclidean space 
with $d=2$ (or $d=3$).  

$\bullet$ Equation (\ref{eq:Euclidean}) holds for a fractal
space with non-integer $d$, in this case $d$ is 
the spectral dimension \cite{stauffer}. 

$\bullet$ It is important 
for us to consider a random walk in a bounded region, because 
the set of conformations is always finite for lattice models, and thus, 
the region of interest in the conformation 
space is also finite.  For a random walk in a bounded ``cavity'' in Euclidean 
space, or on a bounded fractal,  
\be
P(T) \simeq \left\{ \begin{array}{rll}
\left( \frac{2 \pi}{d} T \right)^{-d/2} & {\rm when} & T < T^{\ast} \sim R^2 \\
\frac{1}{{\cal M}} \frac{z_0}{\overline{z}} & {\rm when} & T > T^{\ast} \sim R^2
\end{array} \right. \ .
\label{eq:bounded}
\ee
Here $R$ is the ``size'' of the allowed conformation space (graph), 
${\cal M}$ is the total number of allowed conformations (or graph nodes), 
$z_0$ and $\overline{z}$ are the numbers of possible Monte Carlo moves 
(or incident graph edges), respectively, for the native state and averaged 
over all states.  Equation (\ref{eq:bounded}) means that 
a random walk does not feel the bounds at ``small'' times until 
it arrives at the boundary.  At later times, 
it covers all of the available region in a uniform manner, then the 
probability for visiting each point (conformation), $\alpha$, is 
simply proportional to the number 
of ways, $z_{\alpha}$, incident to that point.  To better understand
the meaning of $R$, it is useful to define the ``distance'' 
$R_{\alpha \beta}$ between two conformations, $\alpha$ and $\beta$, as the 
minimal number of elementary moves necessary to convert 
$\alpha$ into $\beta$.  Then, according to graph theory, 
the diameter of the graph representing our conformation 
space should be defined as the maximum of $R_{\alpha \beta}$ 
over all pairs of conformations.  Our $R$ is then typically on the 
order of one half of this diameter.  Note that in the 
limit of very long loops, $N \gg 1$, we expect the space diameter 
and $R$ to scale as $R \sim N$.    

$\bullet$ Equation (\ref{eq:bounded}) represents a particular example 
of switching, or crossing over, from one dimension to the other.  
In other words, the conformation space may appear to have one dimension 
close to the native state and another 
dimension far from the native state.  In this sense, a bounded space
has the dimension $d$ at small scales and dimension $0$ (like 
a point) at larger scales.   
 
$\bullet$ For a random walk in a $d$-dimensional Lobachevskii space 
\cite{Novikov,Gruet}
\be
P(T) \sim  T^{-d/2}  \exp \left( - T \lambda / 2d \right)  \ ,
\label{eq:Lobachevskii}
\ee
where $\lambda$ is the Gaussian curvature (inverse squared curvature radius) 
of the space.  This formula can be explained in the following simple way.  
At the scale $T \ll 1/\lambda$, when typical distance from the origin 
remains smaller than the curvature radius, 
the space appears effectively flat and equation (\ref{eq:Lobachevskii}) reduces
to (\ref{eq:Euclidean}).  On the other hand, for very
large $T$, $P(T)$ is dominated by the exponential term, which can be 
understood if one remembers that a Cayley tree graph is the discrete counterpart 
of Lobachevskii space.  
It is important to consider Lobachevskii 
geometry because, as we mentioned, the conformation space diameter 
scales as $N$ for very long loops in the $N \gg 1$ limit, 
while the number of conformations scales exponentially with $N$.  
This means that there is an exponential growth of the number of conformations 
as a function of the distance from any given conformation (e.g., native).  
Such an exponential growth is the signature of Cayley tree or Lobachevskii geometry 
\cite{nechaev}.   

$\bullet$ The analysis of loop conformations which arise from a fixed nucleus
of contacts becomes more complicated if we have 
multiple loops.  Still, if loops are independent of one another, then
a simple estimate for the 
conformational space of all the loops combined can be 
obtained.  Consider $k$ loops each having a fraction $f_i$ of the total
number of movable monomers.  When a monomer is chosen for a 
Monte Carlo move, the probability that it will be 
from loop $i$ is $f_i$.  Assume that each loop lives in an 
unbounded Euclidean space so that the probability for loop $i$ to
fold after $t_i$ Monte Carlo steps is $P(t_i)=t_i^{-d_i/2}$, 
neglecting constant factors. The probability for all
loops to return after time $T$, $P(T)$, is thus
\be
P(T)=\sum_{t_1,...,t_k=0}^{T} \delta \left(T - \sum_{i} t_i \right)  
\prod_{i=1}^{k} f_{i}^{t_i}P_i(t_i) \frac{T!}{\prod
t_i!} \ .
\ee
Using (i) Stirling's approximation, (ii) $P_i(t)\sim t^{-d_i/2}$, 
and (iii) noticing that due to the 
combinatorial factor, the sum is dominated at large $T$
by the term in which $t_i=f_iT$, we obtain
\begin{eqnarray}
P(T) & \simeq & \frac{T!}{\prod_{i=1}^k t_i!}\prod_{i=1}^{k}f_i^{f_iT} 
(f_iT)^{-d_{i}/2} \nonumber \\
& \sim & T^{-d_{\rm eff}/2} \ , \mbox{\ \   where \ \ \  } d_{\rm eff}=\sum_{i=1}^{k}d_i 
\ .
\label{eq:deff}
\end{eqnarray}
For independent loops dimensions simply add to each other.  

$\bullet$ In reality, different loops are not independent.  To some 
extent they obstruct each other's folding.  In general, for obstructing 
loops, we expect
\be
d_{\rm eff} \leq \sum_{i=1}^{k}d_i  \ .
\ee

With the summary of mathematical results for the return probability 
in different geometries, we can now proceed to the computer 
experiments \cite{restricted_Monte_Carlo}  
on the loops shown in Figure \ref{fig:loops}.  

\begin{figure}
\centerline{\epsplace{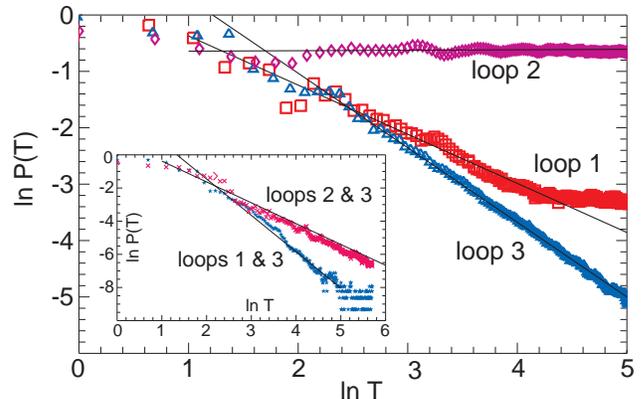}}
\caption{
Log-log plot of the return probability as a function of return time for loop 1,
loop 2, and loop 3.  The dimensions in conformation space for loops 1, 2, and 3 are
$d_1=1.74$, $d_2=0$, and $d_3=2.64$.  The lines
shown have slopes of $-d_i/2$.  
{\bf Inset:} Probability of return for loops 1\&3, and  for loops 2\&3 combined.  
The corresponding dimensions are $d_{13}=4.40 \approx d_1+d_3$
and $d_{23}=2.52 \approx d_2 + d_3$.}
\label{fig:shortloops}
\end{figure}

The return probability for loops 1, 2 and 3 is shown in 
Figure~\ref{fig:shortloops} as a function of return time.  
The log-log graphs indicate clearly the power law dependence 
characteristic of Euclidean geometry (\ref{eq:Euclidean}).  
The dimensions in loop space for loops 1, 2, and 3, 
according to the Figure~\ref{fig:shortloops}, are 
$d_1=1.74$, $d_2=0$, and $d_3=2.64$, respectively.

To see why 
the dimension for loop 2 is 0, note that there are only two
possible positions for loop 2.  At any given time
the probability for loop 2 to be in one position or the other
is 1/2, which means that the probability of 
return must be $P(T)=1/2$ or $\ln P(T)=-0.69$, 
consistent with the result shown 
in Figure~\ref{fig:shortloops}.  
The leveling off expected according to equation 
(\ref{eq:bounded}) in a bounded space is also seen 
for loop 1.  The saturation level (which  
corresponds to $1/P \approx \exp (3.3) \approx 27$) 
and saturation time ($T^{\ast} \approx 67$) are roughly 
consistent with both first and second line of the equation 
(\ref{eq:bounded}) given that the number of conformations 
for loop 1 is ${\cal M} = 51$, and the number of allowed moves 
for the native state is $z_0 =4$.
Since loop 3 is much longer, it 
will undoubtedly exhibit leveling off, but at longer times, 
which we did not reach in our Monte Carlo experiment 
\cite{short_time_comment}.  

To see the effect of having multiple loops on the
loop space dimension, we examine
conformations in which both loops 1 and 3 are allowed to move and those 
in which loops 2 and 3 are allowed to move.  The reason for this choice
of loops is that loops 1 and 3 (and loops 2 and 3) are sufficiently
far apart on the conformation that their interactions
are negligible, consistent with our simple analytical estimate.  
The effective combined dimension for
loops 1 and 3 is $d_{13}=4.40 \approx d_1+d_3$ and for loops 2 and 3 is 
$d_{23}=2.52 \approx d_2 + d_3$ (see Figure~\ref{fig:shortloops})
in agreement with the approximations given in Equation (\ref{eq:deff}).

\begin{figure}
\centerline{\epsplace{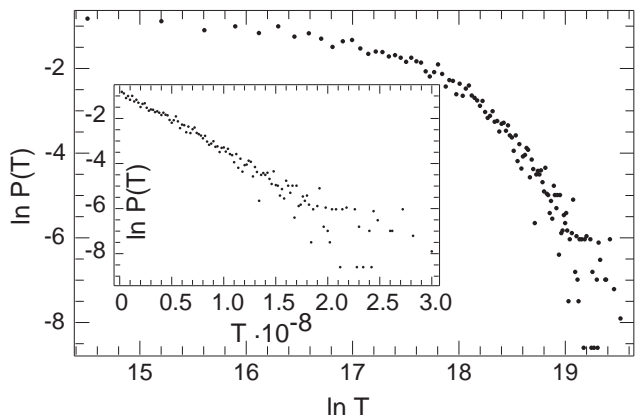 }}
\caption{
Return probability as a function of return time for loop 4.  
Since the return time is typically very long, it is difficult to 
gather statistics.  Thus, the data were binned over the time intervals of 
$2 \times 10^6$, which is why the log-log plot is not linear at 
small times.  The {\bf inset} shows the same data in semi-log scale 
and indicates exponential behavior at long times, which is 
consistent with Lobachevskii geometry.  The least square fit    
yields $d = 1.5$ and $\lambda = 4.9 \times 10^{-8}$.}
\label{fig:longloop}
\end{figure}

The conformational space becomes even more interesting with 
longer loops such as loop 4 (see 
Figure~\ref{fig:loops}), which goes from monomer 20 to 33.  
As Figure~\ref{fig:longloop} indicates, the behavior of 
$P(T)$ for this loop is consistent with
Lobachevskii geometry in which the return probability 
follows a power law at small return times but 
decays exponentially at sufficiently large return times
(see equation (\ref{eq:Lobachevskii})).  A least-squares fit of
$P(T)$ gives $d_4 = 1.5$, and $\lambda = 4.9 \times 10^{-8}$.
Thus, at rather small scales the conformational space of the loop 4 
is a usual fractal graph, while at larger scales it branches  
exponentially like a Cayley tree.  
Physical nature of branchings in the conformation space is very 
simple:  when two different pieces of polymer are close together, each 
piece can move either on one or on the other side of the second piece, 
and to switch from one side to the other it has to go back, which is 
precisely the description of bifurcation point on the Cayley tree. 

To conclude, simple Monte Carlo techniques  
are sufficient for obtaining dimensions of loops. 
In particular, we have shown that there exists 
a nontrivial geometry of the conformational space, with 
noninteger dimensions and with Lobachevskii type 
curvature.  Returning to the introduction and the relation 
between Levinthal's paradox and random walks in 
conformation space, one can ask:  how long 
does it take for a random walk to bring the system 
into the critical region $\omega$?  The most naive 
estimate of this time, $t$, would be $t \sim \left. \left| 
\Omega \right| \right/ \left| \omega \right|$, where 
$\Omega$ is the entire conformation space, and $\left| \ldots \right|$ 
means the number of conformations in the domain $\ldots$.  
This estimate is consistent with the original Levinthal 
formulation \cite{Levinthal}, except $\left| \omega \right| = 1$ there.  
However, we now know that such estimates are only valid in the 
long time limit of a walk in a bounded space (second line 
of equation  (\ref{eq:bounded})).  Thus an understanding
of random walks in conformational space is crucial to 
the understanding of protein folding.


The work was supported by NSF grant DMR-9616791.  AG acknowledges 
useful discussion with S.Nechaev.


\vspace{-.6cm}

\begin{thebibliography}{99}
\vspace{-1.5cm}
%
\bibitem{Levinthal} Levinthal, C.  Mossbauer Spectroscopy in Biological Systems, ed. Debrunner, P., Tsibris, J. C. M. and Munck, E.  U. of Illinois Press, Urbana, 1968.
%
\bibitem{Fersht} Fersht, A. R. {\it Curr. Opin. Struct. Biol.}, 
{\bf 5}, 79-84 (1995).
%
\bibitem{Onuchic} Betancourt, M. R. and Onuchic, J. N.  {\it J. Chem. Phys.},
{\bf 103}, 773-787 (1995).
%
\bibitem{Wolynes} Bryngelson, J. D., Onuchic, J. N., Socci, N. D., and Wolynes, P. G.  {\it Proteins}, {\bf 21}, 167-195 (1995).
%
\bibitem{Shakhnovich} Abkevich, V. I., Gutin, A. M. and Shakhnovich, E. I. 
{\it J. Mol. Biol.}, {\bf 252}, 460-471 (1995).
%
\bibitem{landscape_note}  It is worth stressing the difference between energy landscape
and free energy landscape.  The former is a function of the huge 
number of microscopic coordinates which define a conformation with all the 
microscopic details.  The latter, on the other hand, involves a certain 
preaveraging, or coarse-graining, and is a function of relatively few 
macrovariables.  Unfortunately, these two concepts are frequently 
confused in the literature.    
%
\bibitem{Shakhnovich_nucleus} Abkevich, V., Gutin, A., and Shakhnovich, E.
{\it Biochemistry}, {\bf 33} 10026-10036 (1994).
%
\bibitem{Thirumalai_nucleus}  Klimov, D. K. and Thirumalai, D. {\it J. Mol. Biol.}, 
{\bf 282}, 471-492 (1998).
%
%
\bibitem{Doi-Edwards} Doi, M.  and Edwards, S. F.  The theory of polymer 
dynamics.  Oxford, Oxford, 1986.
%
\bibitem{Homopolymer_collapse} Halperin, A. and Goldbart, P. M. in preparation.
%
\bibitem{Sfatos} Sfatos, C. D., Gutin, A. M. and Shakhnovich, E. I. {\it Phys. Rev. E}, 
{\bf 48}, 465-75 (1993).
%
\bibitem{RMP} Pande, V. S., Grosberg, A. Yu., Tanaka, T.
{\it Rev. Mod. Phys.}, to be published (1999)
%
\bibitem{Mirny} Mirny, L. A.  Harvard Thesis (1998).
%
\bibitem{Homo_Folding} Du, R., Pande, V. S., Grosberg, A. Yu., Tanaka, T., and Shakhnovich, E. I.  submitted to J. Chem. Phys. 
%
\bibitem{stauffer} Stauffer, D. and Aharony, A. Introduction to Percolation 
Theory.  Taylor \& Francis, London, 1992.
%
\bibitem{Novikov} Novikov, S. P. and Fomenko, A. T.   Basic Elements of Differential 
Geometry and Topology, Kluwer Academic, Dordrecht, 1990.
%
\bibitem{Gruet} Gruet, J. C.  {\it Stoch. Stat. Rep.}, {\bf 56}, 53-61 (1996).
%
\bibitem{nechaev} Nechaev, S. Statistics of knots and entangled 
random walks, (World Scientific: Singapore, 1996)
%
\bibitem{nucleus_debate_Wolynes} Wolynes, P. G.  {\it Folding \& Design},
{\bf 3}, R107 (1998).
%
\bibitem{nucleus_debate_Shakhnovich} Shakhnovich, E. I. {\it Folding \& Design}, {\bf 3}, R108-R111 (1998).
%
\bibitem{nucleus_debate_Thirumalai} Thirumalai, D. and Klimov, D. K. 
{\it Folding \& Design}, {\bf 3}, R112-R118 (1998).
%
\bibitem{PGT_current_opinion}  Pande, V. S., Grosberg, A. Yu., Tanaka, T. 
and Rokhsar, D. S.  {\it Curr. Opin. Struct. Biol.}, {\bf 8}, 68-79 (1998). 
%
\bibitem{Pande_Rokhsar} Pande, V. S. and Rokhsar, D. S.  {\it Biochemistry}, 
{\bf 95}, 1490-1494 (1998).
%
\bibitem{restricted_Monte_Carlo} In examining the return probability of 
loops, the usual Monte Carlo procedure
of randomly choosing a monomer to be moved needs to be modified   
such that only the monomers belonging to the loops considered  
should be taken into account.
%
\bibitem{short_time_comment} For very short return times, the return probabilities
do not follow the $T^{-d/2}$ power law simply because the probability at 
zero time is not  infinity.
%
\end{thebibliography}
\end{document}